\documentclass{raa}           
\usepackage{graphicx,times}
\usepackage{natbib}
\usepackage{amssymb,amsmath}
\bibpunct{(}{)}{;}{a}{}{,}

\usepackage[a4paper=true,dvipdfm=true,pagebackref=true]{hyperref}
\hypersetup{pdftitle = The title of my PDF, pdfauthor = My name, pdfsubject= The subject, pdfkeywords = keyword1 keyword2 keyword3} 
\hypersetup{colorlinks = true, linkcolor = green, anchorcolor = red, citecolor = blue, filecolor = red, pagecolor = red, urlcolor = red}

\begin{document}

   \title{Drag-driven instability of a dust layer in a magnetized protoplanetary disc}

% \volnopage{ {\bf 2012} Vol.\ {\bf X} No. {\bf XX}, 000--000}
   \setcounter{page}{1}

   \author{Mohsen Shadmehri\inst{1}, Razieh Oudi\inst{2}, Gohar Rastegarzade \inst{2}}

   \institute{ Department of Physics, Faculty of Sciences, Golestan University, Gorgan 49138-15739, Iran \\
        \and
             Department of Physics,Semnan University, Semnan 35196-45399, Iran\\}
	%\and
%	  Center for Astrophysics, University of Science and Technology of China, Hefei 230026, China\\
%Key Laboratory for Research in Galaxies and Cosmology, The University of Science
%and Technology of China, Chinese Academy of Sciences, Hefei, Anhui, 230026, China\\
%\and 
%Polar Research Institute of China,
%Jinqiao Rd. 451, Shanghai, 200136, China\\
\vs \no
   %{\small Received 2012 June 12; accepted 2012 July 27}
%}

\abstract{ We study drag-driven instability in a protoplanetary disc consisting of a layer of single-sized dust particles which are coupled to the magnetized gas aerodynamically and the particle-to-gas feedback is included. We find a dispersion relation  for axisymmetric linear disturbances and growth rate of the unstable modes are calculated numerically. While the secular gravitational instability in the absence of  particle-to-gas feedback predicts the dust layer is unstable, magnetic fields significantly amplifies the instability if the Toomre parameter for the gas component is fixed.   We also show that even a weak magnetic field is able to amplify  the instability  more or less irrespective of the  dust-gas coupling.}


   \authorrunning{M. Shadmehri, R. Oudi, G. Rastegarzade}            %author_head in even pages
   \titlerunning{Drag-driven instability}  % title_head in odd pages
   \maketitle

%________________________________________________ sections below
%
\section{Introduction}           %% first-level sections will be auto-capitalized

%\label{s:?}
%

%
While the outer parts of protoplanetary discs are prone to the gravitational instability, the inner parts are stable to the gravitational perturbations \citep[e.g.,][]{rafikov2005}. It is known that the onset of the gravitational instability in an accretion disc occurs when the Toomre parameters becomes less than a threshold value around unity and the survival of the newly formed fragments is guaranteed when the cooling time-scale is less than a few dynamical time-scale \citep[e.g.,][]{gammie2001}. Although the {\it dissipationless} gravitational instability  is able to explain some of the observational features of structure formation in the  protoplanetary discs \citep[e.g.,][]{matzner2005}, presence of the dust particles can introduce new physical mechanisms in order dust particles clump  together to form larger objects that may eventually  growth into planet embryos \citep[e.g.,][]{chiang2010}.  This is mainly  because of  the interactions between dust particles and the gas.  Drag force is  proportional to the relative velocity of dust and gas components. But the effect of this exchange of momentum is much stronger on the dust component simply because mass of gas is much larger than the total mass of dust particles.   Dynamics of dust particles in a protoplanetray disc is not necessarily the same as gas component. They rotate slower than the local Keplerian velocity because of the pressure gradient which acts opposite to the direction of the central gravitational force. But an individual particle does not accelerate by the pressure gradient when its internal density is much larger than gas density and thereby dust particles rotate at full Keplerian velocity.

Although a few authors had already studied gravitational stability of accretion discs consisting of dust particles and gas \citep[e.g.,][]{coradini81,noh91}, during recent years specific types of instabilities have been identified for clumping of dust particles   in the protoplanetary discs  which are actually driven by the movement of dust particles through the gas \citep[e.g.,][]{youdin2005,youdin2007,Jac,armit,Laibe} or dust-gas interaction \citep[e.g.,][]{sekiya83,Cuzzi,youdin2011}.  Streaming instability has been studied by many authors during recent years in the linear regime and its non-linear evolution investigated via direct numerical simulations.

Drag driven instability is known as secular gravitational instability which is actually the dissipative version of the classical gravitational instability for a dust layer in a fixed background gas component. Dynamics of the dust particles is mostly affected by gas-dust friction driven instabilities. Irrespective of the strength of the self-gravity, this instability is unconditional and can give raise to clumping of dust particles. There are  simple theoretical explanations for this trend as have been clarified by \cite{goodman2001} and \cite{Cuzzi}. Radial perturbation leads to concentric rings of dust particles with slightly larger density comparing to their ambient dust density. Particles at the outer edge of a ring feel larger gravitational force due to the accumulated mass of the ring and thereby will rotate faster. Since the drag force is proportional to the velocity, inflow of dust particles increases at the outer edge of the ring. At the inner edge, the particles are orbiting at less than Keplerian velocity because of the extra outward gravitational force. Thus,  the particles will therefore be energized by gas drag and will drift toward the ring. No matter how much self-gravity is weak, the mentioned process will eventually give raise to clumping of dust particles.

Most of the previous linear studies of secular gravitational instability assume that dust particles are moving in a fixed background gaseous component \citep[e.g.,][]{Cuzzi,youdin2011,Mich,shadmehri16}. In these models, dust grains are treated as {\it pressure-less fluid}. The nondimensional gas friction time or dimensionless stopping time which is defined as the product of the gas friction time and the Keplerian angular velocity determines gas-dust coupling via the drag force. When dimensionless stopping time is greater than unity, dust particles are decoupled from the gas component and it would not adequate to describe their dynamic using fluid approximation \citep[e.g.,][]{Jalali}.
  
Neglecting gas dynamics is justified by the fact that the total mass of dust particles is much smaller than the mass of the gaseous component of disc and so only dynamics of dust particles is modified because of drag force. Although this argument seems to be reasonable, just recently \cite{Taka} showed that  long-wavelength perturbations are stable when  the dynamical feedback from dust grains in the gas component is considered. Their analysis implies that  we can not neglect small terms in the equation of motion for small  growth rates. Thus, any physical agent that can modify gas dynamics may also affect dust dynamics indirectly via the drag force. Considering the important role of magnetic fields in the structure of protoplanetary discs, it is our motivation to study gravitational instability of a dust layer in a {\it magnetized} gaseous disc which has not been studied before to the best of our knowledge.

Structure of a protoplanetary disc strongly depends on the level of ionization and magnetic fields.  External ionization sources such as X-ray radiation from the central star and cosmic rays can efficiently ionize  surface layers  of a disc.  Most regions of a protoplanetary disc (PPD) are weakly ionized, however, which implies that the coupling between the disc material and the magnetic field to be incomplete. This will eventually lead to the non-ideal MHD effects which appear because of the drift velocity between neutral particles and ionized species. There are three non-ideal MHD effects, i.e. the Ohmic resistivity, Hall effect, and ambipolar diffusion. When the density is high and the ionization is very low,  the Ohmic term is dominant, but the ambipolar diffusion  term influences in the opposite limit. In between these extreme cases, the Hall term plays a significant role. All these non-ideal terms not only significantly modify growth rate of the magnetorotational instability and its non-linear evolution, but also dynamical structure of the disc and launching of winds and outflows are affected by these effects. In this study, we neglect possible role of the non-ideal effects for simplicity.  An important mechanism for transporting angular momentum in an accretion disc which leads to accretion is known as magnetorotational instability  and operates  in weakly ionized discs \citep{balbus91}. Magnetic fields may also provide an efficient mechanism for launching jets or outflows from a disc. Moreover, dynamical structure of a disc is significantly modified in the presence of magnetic fields. Gravitational stability of an accretion disc in the presence of magnetic field has also been studied by many authors \citep[e.g.,][]{Elme,gamm,Fan97,Lizano2010, Lin14}. Many of the previous studies concentrated on analyzing gravitational stability of purely gaseous discs and do not consider dynamics of the dust particles explicitly. \cite{Lizano2010} extended the classical Toomre criterion to a magnetized disc by introducing a modified Toomre parameter which should be greater than one for a gravitationally stable disc. They showed that magnetic tension and pressure stabilize the disc against axisymmetric gravitational perturbations which means magnetic fields suppress  gravitational instability in the protoplanetary discs.

In our study, we consider a disc consisting of the magnetized gas and dust where they are coupled via drag force  and the particle-to-gas feedback is included. We then explore possible effects of the magnetic fields on gravitational stability of the dust layer using a linear perturbations analysis. In the next section, main assumptions and the basic equations of the model are presented. Linearized equations and the resulting dispersion relation are obtained in section 3. Numerical analysis of the unstable modes and their dependence on the input parameters including strength of the magnetic field are presented in section 4. We conclude with a summary of the results. 
 \section{General Formulation}
%

%% Math 

We consider a protoplanetray disc around a central star with mass $M$ as a system consisting of gas and dust components with the momentum exchange. It is assumed that the disc is so thin  that the motion of both gas and dust fluids are in the plane of the disc. It means that we do not consider vertical motion of dust particles.  Previous linear studies of drag-driven instability in a dust layer have been done in the  shearing sheet approximation \citep{gold}. Here, we do not follow this approach. Our linear analysis is performed in cylindrical coordinates $(r,\phi , z)$ where the central star locates at its origin and time-evolution of the perturbations with wavelengths much smaller than the radial distance (i.e., WKB approximation) is studied. Our basic equations for the gas component is similar to \cite{Lizano2010} who studied gravitational stability of a thin and magnetized accretion disc. But we include the drag force due to the interaction with the dust fluid. Since we assume the dust particles are neutral, they do not feel magnetic force.

Thus, basic equations for the gas component  are
\begin{equation}
\frac{\partial \Sigma}{\partial t}+{\nabla}.(\Sigma\mathbf{w})=0 ,
\end{equation}
\begin{displaymath}
\Sigma(\frac{\partial\mathbf{w}}{\partial t}+\mathbf{w}.\nabla\mathbf{w})=
\end{displaymath}
\begin{equation}
-\Sigma \nabla (\Phi-\frac{GM}{r})-c_s^2\nabla\Sigma+\frac{1}{4\pi} \int \mathbf{J}\times\mathbf{B} dz+\frac{\Sigma_d(\mathbf{w_d-w)}}{t_{\rm stop}},
\end{equation}
\begin{equation}
\frac{\partial\mathbf{B}}{\partial t}=\nabla\times(\mathbf{w}\times\mathbf{B}),
\end{equation}
\begin{equation}
\nabla.\mathbf{B}=0,
\end{equation}
where $\Sigma$, $\bf{w}$ and $c_s$ are surface density, velocity and the sound speed of gas, respectively. Also, $t_{\rm stop }$ is the stopping time (see below for its definition). It is assumed that the gas is isothermal.  Magnetic field of gas is denoted by ${\bf B}$ and the current density is ${\bf J} = \nabla \times {\bf B}$. Note that $\Phi$ is the gravitational potential due to both gas and dust fluids. Note that the equations are integrated perpendicular to the disc so that vertically averaged physical quantities do not depend on the vertical coordinate $z$. 

 Also, the basic equations for the dust fluid are written as
\begin{equation}
\frac{\partial\Sigma_d}{\partial t}+\nabla.(\Sigma_d\mathbf{w_d})=D\nabla^2\Sigma_d , 
\end{equation}
\begin{displaymath}
\Sigma_d(\frac{\partial\mathbf{w_d}}{\partial t}+\mathbf{w_d}.\nabla\mathbf{w_d})=
\end{displaymath}
\begin{equation}
-\Sigma_d \nabla (\Phi-\frac{GM}{r})+\frac{\Sigma_d(\mathbf{w-w_d)}}{t_{\rm stop}},
\end{equation}
where ${\bf w_d}$  is dust velocity and   $D$ is the radial diffusivity of the dust component because of the gas turbulence. The diffusion of dust particles due to stochastic forcing by gas turbulence has been studied by many authors \citep[e.g.,][]{youdin2007}. According to Equation (36) of Youdin \& Lithwick (2007), the radial diffusion coefficient $D$ is written as
\begin{equation}
D=\frac{1+\tau +4\tau^2}{(1+\tau^2)^2}D_g,
\end{equation}
 where $D_g$ is is the strength of turbulent diffusion in the gas which can be defined as
\begin{equation}
D_g=\alpha c_s^2 \Omega^{-1},
\end{equation}
where $\alpha$ is the dimensionless measure of turbulent intensity.  Strength  of dust diffusion is measured by the  dimensionless   diffusivity coefficient $\xi$ as $\xi = D/( c_{\rm s}^2 \Omega^{-1})$. Moreover, $\tau$ is the dimensionless stopping time (see below). We note that equation of continuity with the diffusion term is not used commonly. In fact, one can start from the Boltzmann equation to obtain the above hydrodynamical equations which leads to viscosity in equation of motion instead of the diffusion term in equation of continuity. Following previous works \cite[e.g.,][]{Taka} we also used this problematic formulation, although these aspects of the work need further studies.

In the above equations,  $t_{\rm stop }$ is the stopping time which is a time-scale for decay of relative velocity between the gas and the dust due to the drag force. We can then define nondimensional stopping time $\tau = t_{\rm stop} \Omega_{\rm K}$ \citep[e.g.,][]{miyake15}, where angular Keplerian velocity is $\Omega_{\rm K}=\sqrt{GM/r^3}$. If we assume that all dust particles are spherical with the same radius $a$ and homogeneous internal density $\rho_{\rm m}$, then the nondimensional stopping time becomes $\tau = [\rho_{\rm m} a /(\rho_{\rm g} c_{\rm s})] \Omega_{\rm K}$ where $\rho_{\rm g}$ is the gas density. Note that this relation is valid  when the size of the particles is smaller than the mean free path of the gas. For instance, in the minimum mass solar nebula (MMSN) model of \cite{hayashi81} at the radial distances larger than 1 AU from the central star with one solar mass, the mean free path of gas is larger than 1 cm which implies that the above relation for the stopping time is applicable to the particles smaller than this length. Physical properties of the disc and dust distribution depend on the vertical location as well. But we do not consider vertical variation of the physical quantities and one can then evaluate the nondimensional stopping time at the midplane of MMSN \citep{miyake15}:
\begin{equation}
\tau = 1.8 \times 10^{-7} \left(\frac{a}{1 {\rm \mu m}}  \right) \left(\frac{r}{1 {\rm AU}} \right)^{\frac{3}{2}}.
\end{equation}
The internal density of a dust particle is assumed to be $\rho_{\rm m}=2$ g ${\rm cm}^{-3}$ and the surface density and the sound speed obey power-law functions of the radial distance \citep{hayashi81}:
\begin{equation}
\Sigma (r) = 1.7 \times 10^{3} \left( \frac{r}{1 {\rm AU}}\right)^{-\frac{3}{2}} {\rm g} {\rm cm}^{-2},
\end{equation}
\begin{equation}
c_{\rm s} (r) = 1.0\times 10^{5} \left(\frac{r}{1 {\rm AU}} \right)^{-\frac{1}{4}} {\rm cm} {\rm s}^{-1}.
\end{equation}

Note that our study is a {\it local} linear perturbation analysis based on WKB approximation which means that  we do not consider radial dependence of the initial equilibrium state. But the above physical profiles specify how properties of a disc can vary with the radial distance. It can then be used to calculate growth rate of the unstable modes at a certain radial distance. As for the initial magnetic field, we assume the disc is threaded by a net large-scale vertical field $B_{\rm z0}$ so that the ratio of the gas pressure to the magnetic pressure $\beta$ at the midplane of the disc is uniform through the disc. We then have
\begin{equation}
B_{\rm z0} (r) = 590 \left( \frac{\beta}{1000}\right)^{-\frac{1}{2}} \left(\frac{r}{1 {\rm AU}} \right)^{-\frac{13}{8}} {\rm m G}.
\end{equation}

Finally, our system of equations is closed with the Poisson equation for a thin disc which is written as 
\begin{equation}\label{eq:pois}
\nabla^2\Phi=4 \pi G(\Sigma+\Sigma_d)\delta(z).
\end{equation}
Here, gravitational potential $\Phi$ due to both the gas and the dust components is considered. 

\cite{Lizano2010} studied gravitational instability of a gaseous magnetized disc but without dust particles. They vertically averaged all basic equations including the Lorentz term in the equation of motion. We generalize their final  main equations for a magnetized vertically averaged equations to include dust particles and their momentum exchange with the gas component. 

Thus, the continuity  equation for the gas is
\begin{equation}
\frac{\partial\Sigma}{\partial t}+\frac{1}{r}\frac{\partial}{\partial r}(r\Sigma u)+\frac{1}{r}\frac{\partial}{\partial\varphi}(\Sigma v)=0,
\end{equation}
where $u$ and $v$ are the radial and the azimuthal components of gas velocity ${\bf w}$.  The components of radial and azimuthal Lorentz force are:
\begin{displaymath}
\int_{-H}^{H} \mathbf{J}\times\mathbf{B} dz=\int_{-H}^{H} [(B_z\frac{\partial B_r}{\partial z}-B_z\frac{\partial B_z}{\partial r}-\frac{B_\varphi}{r}\frac{\partial (rB_\varphi)}{\partial r}+\frac{B_\varphi}{r}\frac{\partial B_r}{\partial\varphi})\mathbf{e}_r 
\end{displaymath}
\begin{equation}
-(\frac{B_z}{r}\frac{\partial B_z}{\partial\varphi}-B_z\frac{\partial B_\varphi}{\partial z}-\frac{B_r}{r}\frac{\partial(r B_\varphi)}{\partial r}+\frac{B_r}{r}\frac{\partial B_r}{\partial\varphi}) \mathbf{e}_{\varphi} ] dz,
\end{equation}
where $\mathbf{e}_r$ and $\mathbf{e}_{\varphi}$ are unit vectors in the radial and the azimuthal directions, respectively. We assume the toroidal component of the magnetic field is negligible, i.e. $B_\varphi =0$.  This simplifying assumption not only simplifies the main equations, but it prevents emergence of the magnetorotational instability (MRI) modes in our analysis. In order to understand the dynamics in a magnetized disc, we note that MRI has a vital role. However, our purpose is to illustrate and understand the basic mechanism of the SGI with the magnetic field in the absence of MRI modes.  Then the components of the Lorentz force become
\begin{equation}
\frac{1}{4\pi}\int_{-H}^{H} (\mathbf{J}\times\mathbf{B})_r dz=
\frac{1}{4\pi}\int_{-H}^{H} B_z\frac{\partial B^{+}_{r}}{\partial z}dz-\frac{1}{4\pi}\int_{-H}^{H} B_z\frac{\partial B_z}{\partial r}dz,
\end{equation}
and
\begin{equation}
\frac{1}{4\pi}\int_{-H}^{H} (\mathbf{J}\times\mathbf{B})_\varphi dz=-\frac{1}{4\pi}\int_{-H}^{H} (\frac{B_z}{r}\frac{\partial B_z}{\partial \varphi}+\frac{B^{+}_{r}}{r}\frac{\partial B^{+}_{r}}{\partial \varphi})dz.
\end{equation}
By integrating over $z$, we obtain
\begin{equation}
\frac{1}{4\pi}\int_{-H}^{H} (\mathbf{J}\times\mathbf{B})_r dz=\frac{B_z B_r^+}{2\pi}-\frac{H}{4\pi}\frac{\partial B_z^2}{\partial r},
\end{equation}
and
\begin{equation}
\frac{1}{4\pi}\int_{-H}^{H} (\mathbf{J}\times\mathbf{B})_\varphi dz=-\frac{H}{4\pi r}\frac{\partial}{\partial\varphi}(B_r^{+2}+B_z^2).
\end{equation}

Components of equation of motion for the gas are also written as
\begin{displaymath}
\Sigma[\frac{\partial u}{\partial t}+u\frac{\partial u}{\partial r}+\frac{v}{r}\frac{\partial u}{\partial \varphi}-\frac{v^2}{r}]=
\end{displaymath}
\begin{equation}
-c_s^2\frac{\partial\Sigma}{\partial r}-\Sigma\frac{\partial V}{\partial r}+\frac{B_zB_r^+}{2\pi}-\frac{H}{4\pi}\frac{\partial B_z^2}{\partial r}+\frac{\Sigma_d(u_d-u)}{t_{\rm stop}},
\end{equation}
and
\begin{displaymath}
\Sigma[\frac{\partial v}{\partial t}+u\frac{\partial v}{\partial r}+\frac{v}{r}\frac{\partial v}{\partial\varphi}+\frac{uv}{r}]=
\end{displaymath}
\begin{equation}
-\frac{c_s^2}{r}\frac{\partial\Sigma}{\partial\varphi}-\frac{1}{r}\Sigma(\frac{\partial V}{\partial\varphi})-\frac{H}{4\pi r}\frac{\partial}{\partial\varphi}(B_r^{+2}+B_z^2)+\frac{\Sigma_d(v_d-v)}{t_{\rm stop}}.
\end{equation}
The disc scale-height is $H=c_{\rm s}/\Omega_{\rm K}$.  Also, $B_{r}^{+}$ is the radial component of the magnetic field at the surface of the disc. Note that $V$ is the gravitational potential due to the central star and the components of the disc itself, i.e. $V=-GM/r + \Phi $, where $\Phi$ satisfies Poisson's equation (\ref{eq:pois}). 

The induction equation becomes
\begin{equation}
-\frac{\partial B_z}{\partial t}+\frac{1}{r}[\frac{\partial}{\partial r}(r B_z u)+\frac{\partial}{\partial\varphi}(B_z v)]=0,
\end{equation}

Continuity equation for the dust fluid is
\begin{displaymath}
\frac{\partial\Sigma_d}{\partial t}+\frac{1}{r}\frac{\partial}{\partial r}(r\Sigma_d u_d)+\frac{1}{r}\frac{\partial}{\partial\varphi}(\Sigma_d v_d)=
\end{displaymath}
\begin{equation}
D[\frac{1}{r}\frac{\partial}{\partial r}(r\frac{\partial \Sigma_d}{\partial r})+\frac{1}{r^2}\frac{\partial^2\Sigma_d}{\partial\varphi^2}],
\end{equation}
and the components of equation of motion for the dust fluid are
\begin{equation}
\Sigma_d[\frac{\partial u_d}{\partial t}+u_d\frac{\partial u_d}{\partial r}+\frac{v_d}{r}\frac{\partial u_d}{\partial \varphi}-\frac{v_d^2}{r}]=-\Sigma_d\frac{\partial V}{\partial r}+\frac{\Sigma_d(u-u_d)}{t_{\rm stop}},
\end{equation}
and
\begin{equation}
\Sigma_d[\frac{\partial v_d}{\partial t}+u_d\frac{\partial v_d}{\partial r}+\frac{v_d}{r}\frac{\partial v_d}{\partial\varphi}+\frac{u_d v_d}{r}]=-\frac{\Sigma_d}{r}(\frac{\partial V}{\partial\varphi})+\frac{\Sigma_d(v-v_d)}{t_{\rm stop}}.
\end{equation}

%
%
%\begin{equation}
%V=-\frac{2\pi G}{k}(\Sigma+\Sigma_d)
%\end{equation}
%
%

We note that dust particles are assumed to be neutral, and so, they do not experience magnetic force. Moreover, our dusty fluid is pressure-less and for this reason gradient of pressure does not appear in the above equation of motion.

\section{Linear Perturbations}
Having basic MHD equations including dust contributions, we can now perturb all physical quantities around a uniform equilibrium configuration and then investigate their fate  provided that  perturbations are much smaller than the initial state. This kind of linear analysis will lead to a dispersion relation to specify unstable modes and their growth rates. The subscripts 0 and 1 are used to denote the initial state and the perturbed quantities, respectively. Equilibrium states must satisfy continuity, motion and induction equations. We know that initial states are independent of $t$,$\varphi$ and $z$ and the velocities of dust and gas components are assumed to be same initially. We assume that $\Sigma_0$ is independent of $r$. Moreover, we assume that the radial component of the magnetic field at the surface of the disc is negligible for simplicity. Also, the initial vertical component of the magnetic field is considered to be independent of the radial distance.  Then, equilibrium state satisfied all equations automatically except radial component of motion for gas and dust. The zeroth order of radial component of equation of motion for the gas is 
\begin{equation}
\Omega^2r-\frac{c_s^2}{\Sigma_0}\frac{\partial \Sigma_0}{\partial r}-\frac{\partial V_0}{\partial r}+\frac{B_{0r}^+B_{0z}}{2\pi\Sigma_0}-\frac{H}{4\pi\Sigma_0}\frac{\partial B_{0z}^2}{\partial r}=0,
\end{equation}
 which reduces to $\Omega^2 r-\frac{\partial V_0}{\partial r}=0$ subject to our mentioned simplifying assumptions.

Our linear perturbation of a physical quantity $X$ is $X=X_0 + X_1 e^{{\rm i} (\omega t +kr- m \varphi)}$, where $\omega$ is the frequency and $k$ is the radial wavenumber and $m$ is a positive integer for nonaxisymmetric perturbations and $m=0$ for axisymmetric modes. Here, we consider only axisymmetric perturbations.  Following \cite{Lizano2010}, we also make further assumption that $|k|r \gg 1$, which means wavelength of the perturbations is much smaller than the radial distance. 
 
Thus, the linearized  dynamical equations  for axisymmetric modes expand to 
\begin{equation}
\omega\frac{\Sigma_1}{\Sigma_0}+k u_1=0,
\end{equation}
\begin{displaymath}
i\omega u_1-2\Omega v_1+ikc_s^2\frac{\Sigma_1}{\Sigma_0}+
\end{displaymath}
\begin{equation}
ikV_1+i(1+kH)\frac{B_{z0}B_{1z}}{2\pi \Sigma_0}-\frac{Z(u_{1d}-u_1)}{t_{\rm stop}}=0,
\end{equation}
\begin{equation}
i\omega v_1+u_1\frac{\kappa^2}{2\Omega}-\frac{Z(v_{1d}-v_1)}{t_{\rm stop}}=0,
\end{equation}
\begin{equation}
i\omega B_{1z}+ikB_{z0}u_1=0,
\end{equation}
\begin{equation}
(i \omega +Dk^2 )\frac{\Sigma_{1d}}{Z\Sigma_0}+iku_{1d}=0,
\end{equation}
\begin{equation}
i \omega u_{1d}-2\Omega_K v_{1d}+ikV_1-\frac{(u_1-u_{1d})}{t_{\rm stop}}=0,
\end{equation}
\begin{equation}
i \omega v_{1d}+u_{1d}\frac{\kappa^2}{2\Omega_K}-\frac{(v_{1}-v_{1d})}{t_{\rm stop}}=0,
\end{equation}
\begin{equation}
V_1+\frac{2\pi G}{k}(\frac{\Sigma_1}{1+kH}+\frac{\Sigma_{1d}}{1+kH_{d}})=0,
\end{equation}
where $H_{d}$ is the dust scale height $H_d=\sqrt{\frac{\alpha}{\tau}}H$ and $Z$ is the ratio of the dust density to the gas density or disc metallicity  for the initial state, i.e. $Z=\Sigma_{0d}/ \Sigma_0$. Moreover, $\lambda$ is the dimensionless mass-to-flux ratio and is defined as
\begin{equation}
\lambda=\frac{2\pi G^\frac{1}{2}\Sigma_0}{B_{z0}}.
\end{equation}
The additional parameter resulting in our analysis is the magnetically modified Toomre parameter $Q_{\rm M}$, i.e.
\begin{equation}
Q_{\rm M}=\frac{\Theta^\frac{1}{2}c_s \kappa}{\pi G \epsilon \Sigma_0},
\end{equation}
where $\Theta=1+\frac{B_{z0}^2 H}{2\pi c_s^2 \Sigma_0}$ and $\epsilon = 1-\frac{1}{\lambda^2 }$. Also, $\kappa$ is the epicyclic  frequency where in the absence of the magnetic effects it becomes the Keplerian angular velocity. But magnetic forces reduce the epicyclic  frequency, though its exact  value depends on the geometry of the magnetic configuration. \cite{Lizano2010} approximated the epicyclic  frequency as  $\kappa=f\Omega$ where $f$ is a number less than unity. Obviously, we can assume $f\simeq 1$ for the weak magnetic fields.  

If we introduce the nondimensional growth rate and the nondimensional wavenumber as $x= i\omega /\Omega$ and $y = kH$, then we can re-write the above linearized equations:
\begin{equation}
x \frac{\Sigma_1}{\Sigma_0}+i\frac{y}{c_s}u_1=0 ,
\end{equation}
\begin{displaymath}
[x+\frac{Z}{f\tau}+(\frac{2y(1+y)}{\lambda^2 x})(\frac{\Theta^\frac{1}{2}}{Q_M\epsilon})]u_1-2v_1+
\end{displaymath}
\begin{equation}
iyc_s\frac{\Sigma_1}{\Sigma_0}+i\frac{y}{c_s}V_1-\frac{Z}{f\tau}u_{1d}=0 ,
\end{equation}
\begin{equation}
[x+\frac{Z}{f\tau}]v_1+\frac{1}{2}u_1-\frac{Z}{f\tau}v_{1d}=0,
\end{equation}
\begin{equation}
x B_{1z}+i\frac{y}{c_s}B_{z0}u_1=0,
\end{equation}
\begin{equation}
[ x+\xi y^2]\frac{\Sigma_{1d}}{Z\Sigma_0}+i\frac{y}{c_s}u_{1d}=0,  
\end{equation}
\begin{equation}
[x+\frac{1}{f\tau}]u_{1d}-\frac{2}{f} v_{1d}+i\frac{y}{c_s}V_1-\frac{u_1}{f\tau}=0,
\end{equation}
\begin{equation}
[x+\frac{1}{f\tau}]v_{1d}+\frac{f}{2}u_{1d}-\frac{v_1}{f\tau}=0,
\end{equation}
\begin{equation}
V_1+\frac{2\Theta^\frac{1}{2}c_s^2 }{Q_M \epsilon \Sigma_0 y}(\frac{\Sigma_1}{1+y}+\frac{\Sigma_{1d}}{1+\sqrt{\frac{\alpha}{\tau}}y})=0.
\end{equation}
Thus, we have eight equations and eight unknowns, i.e. $\Sigma_{1}$, $\Sigma_{\rm 1d}$, $v_1$, $u_1$, $v_{\rm 1d}$, $u_{\rm 1d}$, $V_1$, $B_{\rm 1z}$. Since the above linearized equations are valid for the perturbations with a wavelength much smaller than radial distance, i.e. $kr\gg 1$, then we can consider perturbations which satisfy this inequality: $kH=y\gg H/r$. For instance, in a thin disc with $H/r = 0.1$,  we consider only perturbations which are larger than this value, i.e. $y\gg 0.1$.  In addition to this constraint of local approximation, the validity of the vertically integrated equations requires $k u<\Omega_{\rm K}$ \citep[e.g.,][]{wu,kato}. this requirement can be written as $kH<(\alpha H/r)^{-1}$ where $\alpha$ is the disc viscosity. Thus, our analysis is valid for $(H/r)\ll y < (\alpha H/r)^{-1}$. If we set $\alpha =0.01$ and $H/r =0.1$, then the valid range of nondimensional wavenumber becomes $0.1\ll y <10^3$. Also, validity of vertical integrated equations implies that the growth rates of the unstable modes are less than angular velocity \citep{kato}. This requirement  is justified by the unstable modes as we will show.  

Existence of a set of nontrivial solutions for the above linearized equations imply that the determinant of the coefficients becomes zero which gives us an algebraic equation involving the input parameters, growth rate and the wavenumber of the perturbations. Using MAPLE software, we found the dispersion relation. But the equation is very lengthy, and so, we do not bring it here. However, our analysis is based on the roots of this equation which can be calculated numerically. Obviously, unstable modes correspond to the roots with positive real part, i.e. ${\rm Re}(x) >0$. We generally found one or two unstable modes for a given set of the input parameters.

It is useful to re-write the input parameters as follows \citep{Lizano2010}:
\begin{equation}\label{eq:lambda}
\lambda=2.71\mu(\frac{N_{\rm H}}{10^{24} {\rm cm}^2})(\frac{B_{z0}}{\rm 1 mG})^{-1},
\end{equation}
\begin{equation}\label{eq:Theta}
\Theta=1+1.15\times 10^{-2}(\frac{B_{z0}}{\rm 1 mG})^2(\frac{H}{\rm 1 AU})(\frac{N_H}{10^{24} {\rm cm}^2})^{-1}(\frac{T}{\rm 1 K})^{-1},
\end{equation}
\begin{equation}\label{eq:epsilon}
\epsilon=1-1.36\times 10^{-1}(\frac{1}{\mu^2})(\frac{B_{z0}}{\rm mG})^2(\frac{N_{\rm H}}{10^{24} {\rm cm}^2})^{-2},
\end{equation}
\begin{equation}\label{eq:QM}
Q_M=2.12(\frac{\Theta^\frac{1}{2}}{\epsilon\mu^\frac{3}{2}})(\frac{\Omega}{10^{-2} {\rm km} {\rm s}^{-1} {\rm AU}^{-1}})(\frac{T}{\rm 1K})^\frac{1}{2}(\frac{N_{\rm H}}{10^{24} {\rm cm}^2})^{-1},
\end{equation}
where $\mu$ is the molecular weight, $N_H$ is the hydrogen column density, $T$ is the gas temperature and $\Omega$ is the angular velocity.
\begin{table}

\begin{center}

\begin{tabular}{c | c | c | c | c | c }

\hline

$B_{z0} {\rm (mG)}$&      $\lambda$&	$\Theta$&	$\epsilon$&	  $f$&	$Q_M$ \\ \hline

0&	 $\infty$& 1& 1& 1& 1.99\\

5&	4.35& 1.05& 0.94& 1& 2.16\\

10& 2.17& 1.21& 0.78& 1& 2.79\\ 

15& 1.45& 1.48& 0.52& 1& 4.9 \\

\end{tabular}

\end{center}

\caption{Input parameters for different values of the initial vertical magnetic field $B_{z0}$.}

\end{table}

\section{Analysis}

\begin{figure}%[tb]
\includegraphics[scale=0.7]{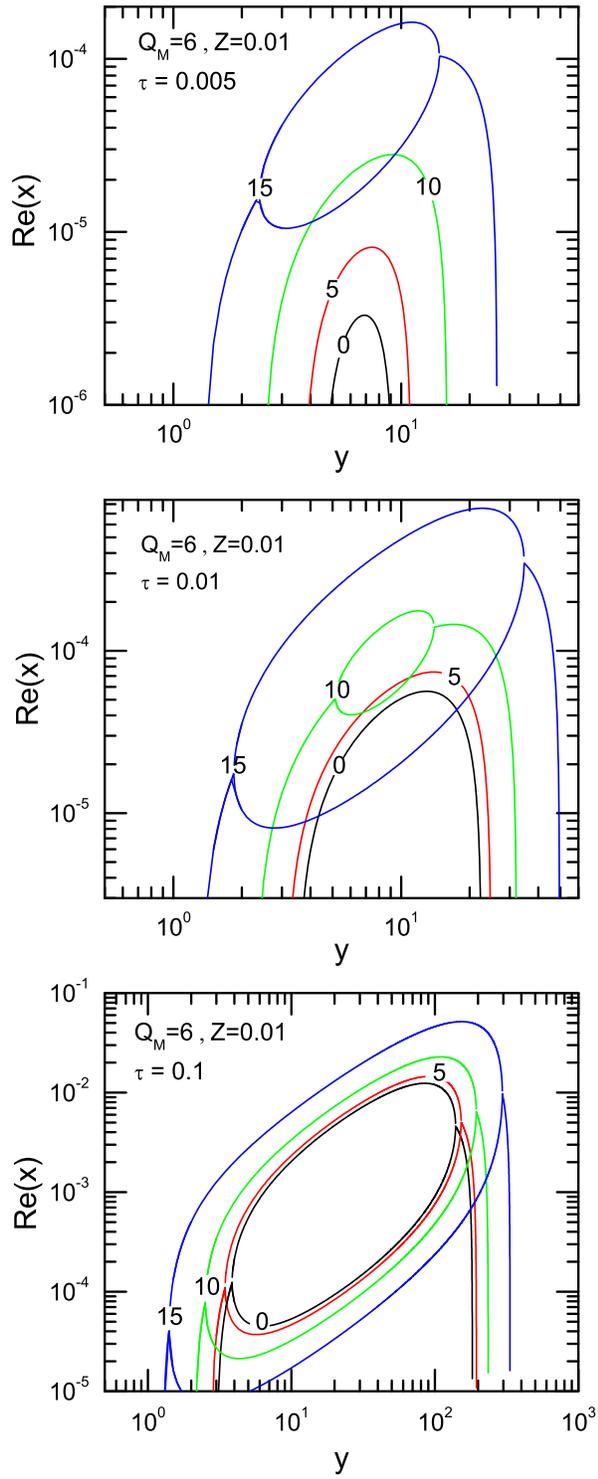}
\caption{ Dispersion relation of the unstable modes for different values of the magnetic strength and dimensionless stopping time. Each curve is labeled by its corresponding value of $B_{z0}$. Here, magnetic Toomre parameter is $Q_{M}=6$ which is larger than the threshold for the instability (see Table 1).} %% no full stop at the end of caption
\label{fig:1}
\end{figure}
\begin{figure}%[tb]
\includegraphics[scale=0.75]{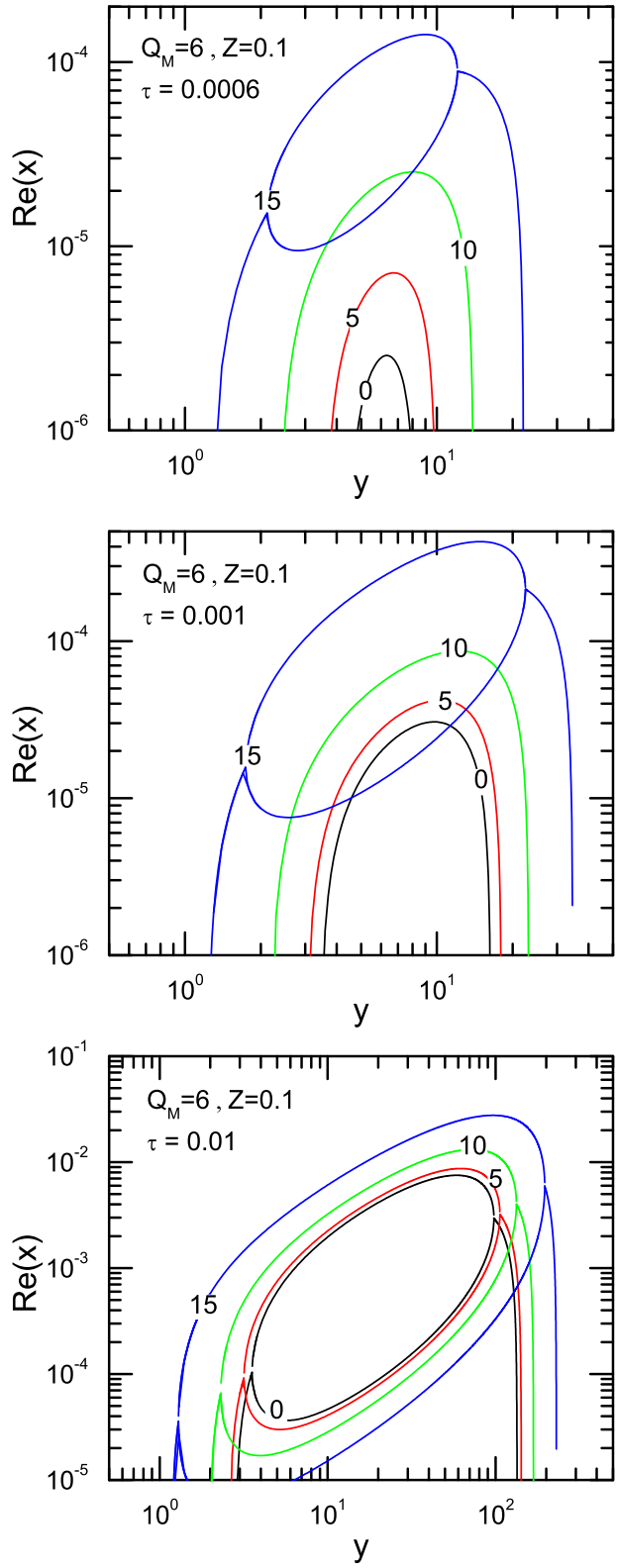}
\caption{ Same as Figure \ref{fig:1}, but surface density of dust particles is larger. } %% no full stop at the end of caption
\label{fig:2}
\end{figure}
%% Math 

%
%
\begin{figure}%[tb]
\includegraphics[scale=0.4]{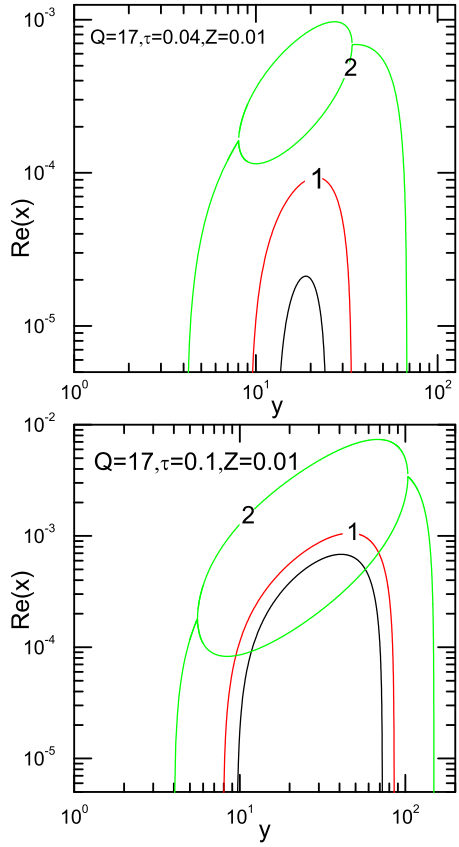}
\caption{ Growth rate of the unstable modes as a function of the normalized wavenumber in the minimum mass solar nebula at the radial distance 100 AU where the Toomre parameter is $Q_M =17$. Other input parameters are same as previous figures, and different values for the dimensionless stopping time $\tau$ are adopted. Each curve is labeled by the corresponding value of the vertical magnetic field and black curve is for the non-magnetized case. We found that for magnetic field strength larger than 2 mG, the parameter $\epsilon$ becomes negative which is not acceptable.} %% no full stop at the end of caption
\label{fig:3}
\end{figure}

We can now investigate axisymmetric unstable modes for different sets of the input parameters in order to explore possible effects of the magnetic field on the drag-driven instability. Nonzero values for $m$  do not affect essentially behavior  of the unstable solutions.   In  Figures 1 and 2, we assume $H = 143 {\rm AU}$, $T=250$ K, $\mu = 2.33$ and $N_{\rm H}=3.46\times 10^{24} {\rm cm}^{-2}$ \citep{Lizano2010}. Corresponding to these input parameters, one can calculate the other parameters based on equations (\ref{eq:lambda})-(\ref{eq:QM}) for different values of the initial vertical magnetic field (Table 1).  We first examine the dependence of the growth rate on the grain size, or dimensionless stopping time. Figure \ref{fig:1} shows unstable growth rate for particles with different sizes, ranging from strongly coupled particles with dimensionless stopping time $\tau =5\times 10^{-3}$ (top panel) and $\tau = 10^{-2}$ (middle panel) to a slightly less coupled case with $\tau = 10^{-1}$ (bottom panel). Each curve is labeled by its corresponding value of $B_{ z0}$. In this figure,  the standard value of disc metallicity is adopted, i.e. $Z=0.01$, and Toomre parameter is $Q_M =6$ and $\alpha=10^{-6}$. Here, different values of $\tau$ are considered. Note that our adopted value of the Toomre parameter is greater than the threshold of the instability which means that the system is stable in the absence of dust particles.  For some of the input parameters, we found two unstable roots where one root is much smaller than the other one. We actually displayed both roots, though the larger root which specifies the most unstable root is more interesting. We consider different values of the disc metalicity (i.e., $Z=0.01$ and $Z=0.1$) in our analysis to explore possible effects of its variations on the instability. Figure \ref{fig:1} shows that the  instability occurs in the presence of the magnetic fields, and as the strength of the magnetic field increases the instability grows faster. This feature is understandable by the fact that the critical value of the Toomre parameter for the instability in the absence of the dust particles increase with the magnetic field (see Table 1). Since we assume a fixed value for the Toomre parameter, the system becomes closer to the threshold of the instability with increasing the magnetic field.  Figure \ref{fig:2} is same as Figure \ref{fig:1}, but the disc metallicity is larger, i.e. $Z=0.1$. Again, the system is unstable in the presence of the dust particles. Figure \ref{fig:1} shows that when the disc contains aerodynamically well-coupled dust particles, even weak magnetic fields  can destabilize  the system considerably. For example, in the strongly coupled case (top panel), the system is unstable in the presence of the magnetic field. But as the level of dust-gas coupling reduces, the growth rate increases. Moreover,  wavelength of the most unstable mode increases with the magnetic field strength when the particles are well-coupled to the gas. Note that in all cases, the corresponding modified Toomre parameter $Q_M$ is larger than one (see Table 1).

Figure 3 shows growth rate of the instability in  the minimum mass solar nebula at the radial distance 100 AU where the Toomre parameter is $Q_M =17$. The rest of the input parameters are the same as previous figures. At the radial distance $100 AU$, when we have $\tau=0.04$, the size of the dust particles is $a=222$ $\mu {\rm m}$ and  the most unstable wavelength is around $1 AU$ and the corresponding growth time is $0.13\times 10^6$ years. For $\tau=0.1$,  the size of the dust particles is $a=555$ $\mu {\rm m}$ and the most unstable wavelength and the corresponding growth time become $2.45 AU$ and $ 10^6$ years, respectively.

Cosmic rays and radiation of the central star are the main sources of ionization in a protoplanetary disc. It is known that there is a region in a protoplanetary disc where neither cosmic rays can penetrate to ionize the gas nor the radiation of the central star is able to ionize the gas. This non-ionized region which is magnetically {\it inactive} is called dead zone \citep{Gam}. But interior to the dead zone or beyond that region,  the gaseous component of the disc is magnetically active. Our analysis shows that drag-driven instability is more efficient in the magnetized regions comparing to the regions where magnetic fields does not play a significant role.   Although the numerical values adopted in previous figures are certainly subject to  uncertainties,  our analysis serves as a proof of concept to illustrate the important role  of the magnetic field in  the drag-driven instability in protoplanetary discs.

\section{conclusion}
We surveyed linear instability of a dust layer in a magnetized gaseous disc for the perturbations with wavelengths much small than the radial distance.  One of the interesting findings in this paper is that magnetic field can amplify the instability for even a weak gas-dust coupling. In particular, we showed that for  well-coupled particles, even a weak magnetic field is able to amplify the instability and leads to a completely unstable system. Our study shows that the greatest response for axisymmetric perturbations occurs at large wavelengths. We also found that in the presence of magnetic fields enhancing the disc metallicity promotes the instability because this enhancement leads to stronger self-gravity of particles and slower radial drift.   

Time-scale of drag driven instability should be shorter than radial drift time-scale if the instability is  responsible for the planetesimal formation. Based on this physical constraint, the minimum dust abundance for planetesimal formation via secular gravitational instability has been estimated by \cite{Takeuchi}. Considering the destabilizing role of magnetic field, however, we think this minimum dust abundance is modified if magnetic fields are considered.

\section*{Acknowledgments}
We are  very grateful to  anonymous referee for his/her very useful comments and suggestions which greatly helped us to improve the paper. MS is grateful to Prof. Shu-ichiro  Inutsuka for his useful comments on the early version of this manuscript. 

\bibliographystyle{raa}
\bibliography{reference}

\end{document}